\documentclass[11pt,a4paper]{article}
\pdfoutput=1 

\usepackage{jinstpub}
\usepackage[utf8]{inputenc}
\usepackage{subfig}
\usepackage{siunitx}

\graphicspath{ {pictures/} }

\title{Versatile systems for characterization of large-area silicon pad sensors for highly-granular calorimetry}

\begin{document}

\author[ab1]{P. Dias de Almeida,\note{Corresponding author.}}
\affiliation[a]{CERN, Switzerland}
\affiliation[b]{University of Cantabria, Spain}

\emailAdd{p.almeida@cern.ch}

\abstract{
High-granularity calorimeters utilizing silicon pad sensors as main active material are being designed for the CMS endcap calorimeter upgrade and have been proposed for the electromagnetic calorimeters at CLIC, ILC and FCC-ee. The silicon sensors in such experiments are foreseen to cover a very large area of O(\SI{1000}{\meter^2}). They are typically produced from 6- or 8-inch wafers and consist of a few hundred smaller cells, each with an area of O(0.1 to \SI{1.1}{\centi\metre^2} ). Currently the CMS endcap calorimeter upgrade is in a prototyping phase with the aim of choosing the final sensor design. Flexible systems are needed for quick sensor characterization as close as possible to operating conditions, for testing different prototypes and for quality control during mass production. The ARRAY system consists of an active switching matrix PCB with 512 input channels and a passive probe card specific for each sensor layout prototype. The probe card makes contact with each individual pad through spring-loaded pins. ARRAY is designed to measure the leakage current and capacitance per-cell for different bias voltages, keeping the entire sensor area under bias. The Hexaboard probe card follows a similar principle: readout electronics foreseen to be used in the final CMS detector are mounted on a PCB that makes contact with the sensor through spring-loaded pins. It is used to perform noise and charge collection efficiency measurements with irradiated sensors without the need for module assembly. We present the design of the ARRAY and Hexaboard systems as well as measurements performed on different CMS prototype silicon sensors. We also compare the results with alternative multi-needle setups.
}


\collaboration[c]{on behalf of the CMS collaboration}

\proceeding{3$^{\text{th}}$ Conference on Calorimetry for High Energy Frontier\\
  25 -- 29 November 2019\\
  Fukuoka, Japan}

\maketitle

\section{Introduction}

The High Luminosity Large Hadron Collider (HL-LHC) is expected to start delivering collisions in 2027, with an instantaneous luminosity of \SI{5}{} $-$ \SI{7.5e34}{\cm^{-2}\s^{-1}}, resulting in a tenfold increase in integrated luminosity from the original LHC program \cite{Apollinari:2284929}. 
To cope with the challenges from this upgrade, CMS is planning several improvements to its detectors, one being the replacement of its endcap calorimeters with a High Granularity Calorimeter (HGCAL) \cite{Collaboration:2293646}.

The new calorimeter will use both silicon and plastic scintillator as a detector medium, with silicon covering the regions subjected to higher fluences up to \SI{e16}{n_{eq}/\cm^2}, and the scintillators up to \SI{8e13}{n_{eq}/\cm^2}. 
In total, more than \SI{600}{\meter^2} of silicon will be used in the final detector, corresponding to \SI{6.1}{M} channels, built with roughly \SI{31000}{} 8-inch sensors. 

A description of the sensors can be found in the next section followed by a section dedicated to the description of three different setups: CV/IV, noise and TCT.

\section{Silicon Sensors for the CMS HGCAL Upgrade}

The silicon sensors are hexagonal in shape as to use the most area of each silicon wafer from which they are built, while still fitting together in a tile pattern. Although the final sensors will be made out of 8-inch silicon wafers, in the prototyping phase sensors made out of 6-inch process were also tested.

Three sensor active thicknesses will be used: \SI{120}{}, \SI{200}{} and \SI{300}{\micro\meter}, to compensate the signal degradation at different radiation levels. To cope with differences in sensor capacitance, different granularities are also being considered, with individual cell sizes varying from \SI{0.5}{} to \SI{1.1}{cm^2}, resulting in a total number of $\sim$200 to $\sim$500 cells per sensor.

\begin{figure}[h]
    \centering
    \subfloat[]{{\includegraphics[height=3.5cm]{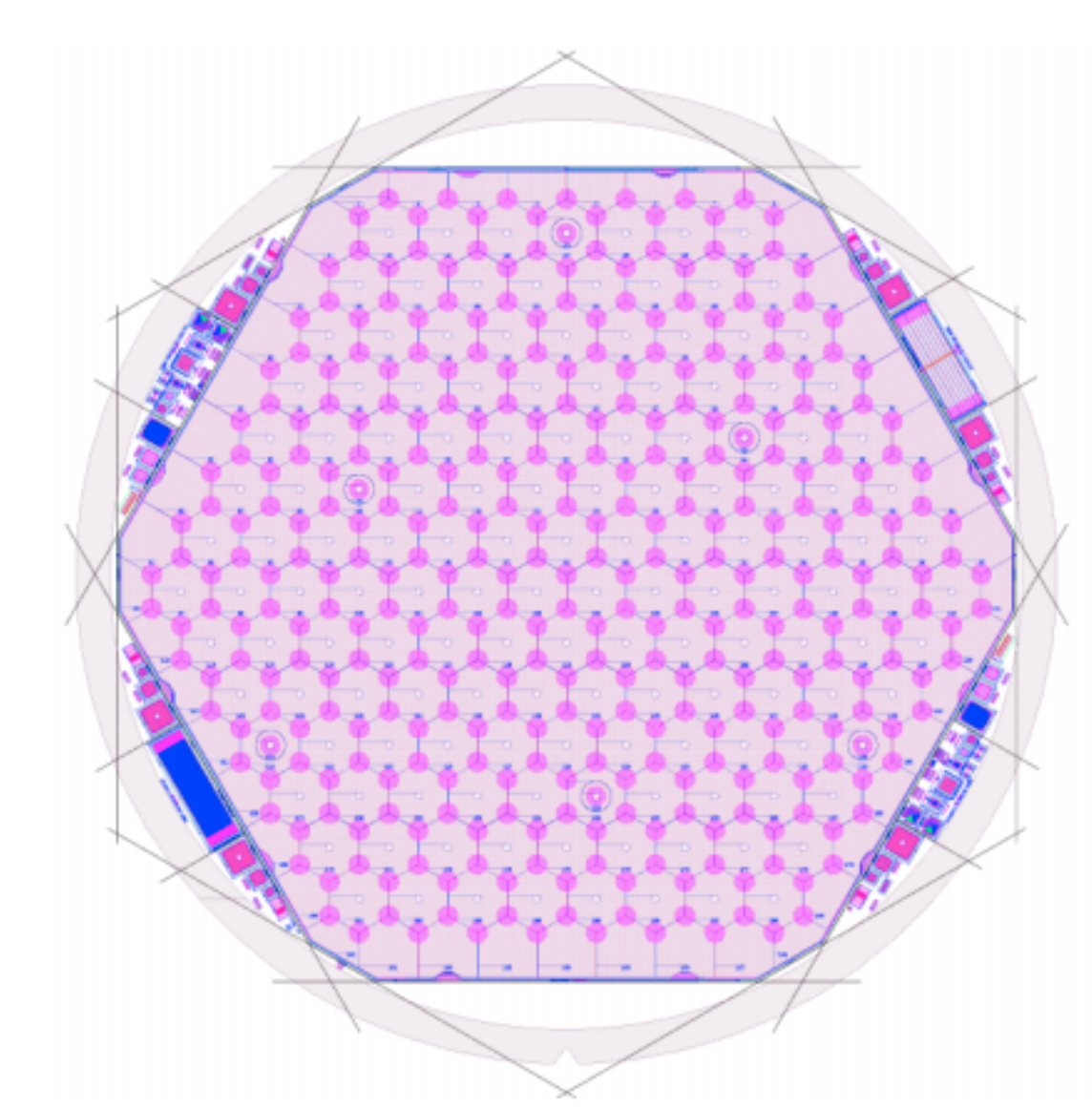} \label{subfig:layout} }}
    \subfloat[]{{\includegraphics[height=3.5cm]{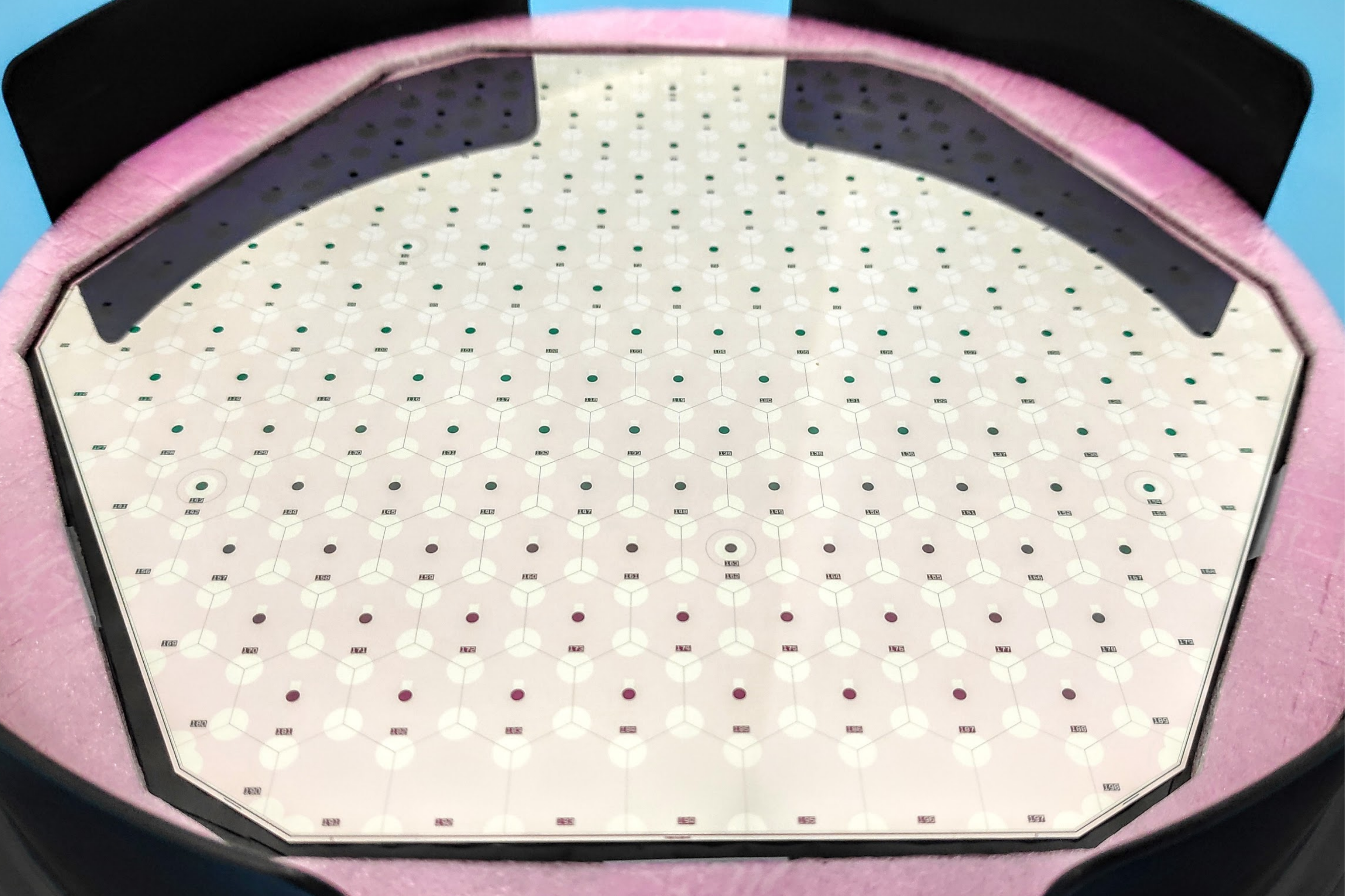} \label{subfig:photo} }}
    \caption{a) The wafer layout of an 8-inch 192-cell prototype sensor, with test structures occupying four of the six leftover regions in the circular silicon wafer not occupied by the hexagonal sensor. b) A picture of the same sensor.
}
    \label{fig:sensor}
\end{figure}

Given the size of the sensors, their large number of cells, and the amount of sensors needed for the final detector, quick and versatile testing methods are needed that operate as close as possible to the operating conditions. 

\section{Versatile Characterization Systems}

\subsection{Capacitance-Voltage (CV) and Current-Voltage (IV) Setup}

\begin{figure}[h]
    \centering
    \subfloat[]{{\includegraphics[height=5cm]{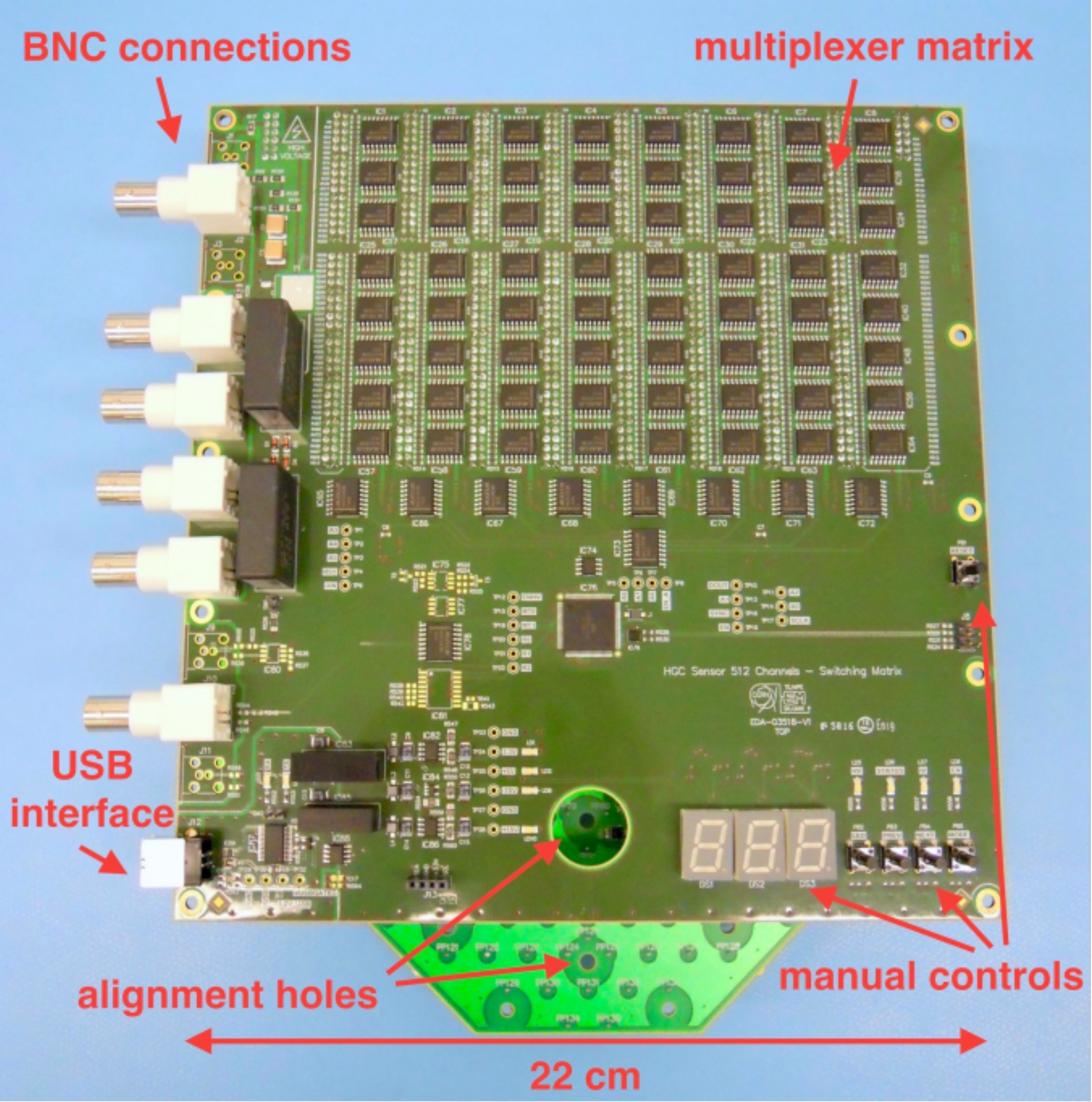} \label{subfig:ARRAY_top} }}
    \subfloat[]{{\includegraphics[height=5cm]{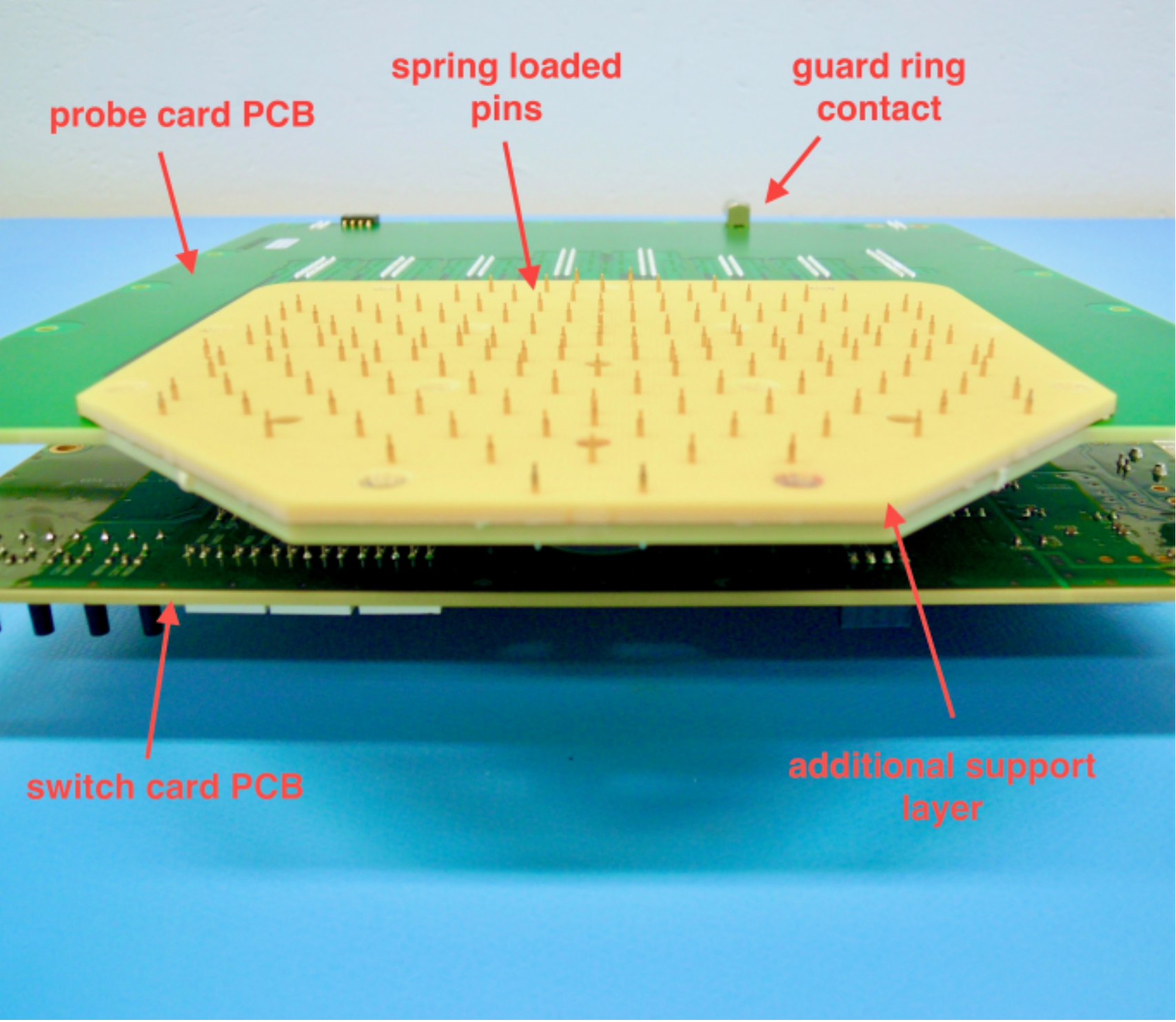} \label{subfig:ARRAY_side} }}
    \caption{Two main components of the ARRAY system \cite{Pitters:2668752}: a) the switch card that contains most of the system's components including connectors, filters, a matrix of multiplexers for selecting the channel and a microcontroller; and b) the probe card that mainly routes the different channels of the switch card to spring loaded pins to be in contact with the sensor.}
    \label{fig:ARRAY_pictures}
\end{figure}

\begin{figure}[h]
    \centering
    \includegraphics[width=0.7\textwidth]{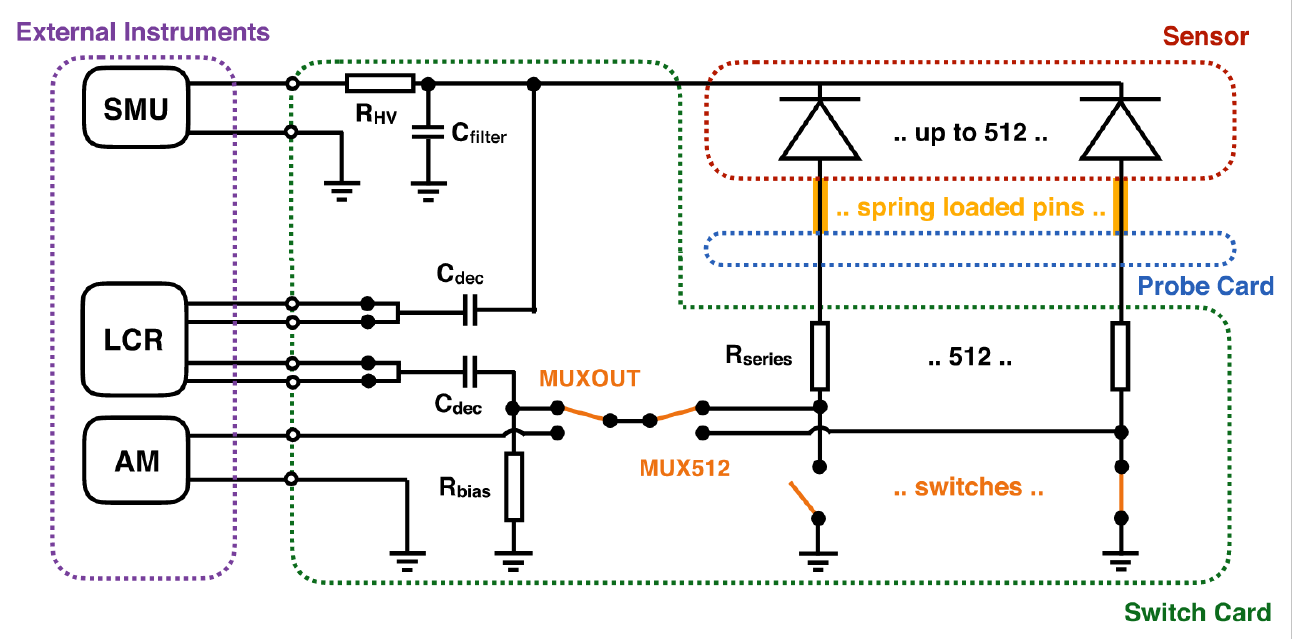}
    \caption{Schematic of the ARRAY system \cite{Pitters:2668752}. The source meter unit (SMU) provides the high voltage, the LCR meter is used to measure capacitance and the ammeter (AM) is used for precise current measurement. $R_{HV}$ and $C_{filter}$ form a low pass filter for the bias voltage, the two $C_{dec}$ are decoupling capacitors used to protect the LCR meter from the bias voltage, $R_{series}$ are resistors in series with each individual channel, protecting the sensor and the sensitive multiplexers, and $R_{bias}$ is needed for not grounding the channel under test when measuring capacitance.} 
    \label{fig:ARRAY_diagram}
\end{figure}

The ARRAY system was developed to measure capacitance and leakage current \cite{Pitters:2668752}. 
These measurements are important to understand properties of the sensor, e.g., power consumption, depletion voltage, effective doping concentration.
The system comprises two PCBs mounted onto each other through pin connectors as shown in Figure \ref{fig:ARRAY_pictures} and a simplified schematic of the system can be seen in Figure \ref{fig:ARRAY_diagram}.

For each sensor layout, a simple probe card with matching pin connectors is used. Keeping the system versatile and easily adaptable.

The spring-loaded pins have a travel distance of \SI{1.4}{\milli\meter}, which allows good contact across the whole sensor even if the probe card and the sensor are not perfectly parallel. 
When compressed, they exert a force of \SI{25}{g} and in general the recommended contact area should be bigger than \SI{1}{mm^2}. It takes around one hour to take an IV or CV measurement of 150 channels with 15 voltage steps.

\begin{figure}[h]
    \centering
    \subfloat[]{{\includegraphics[height=5cm]{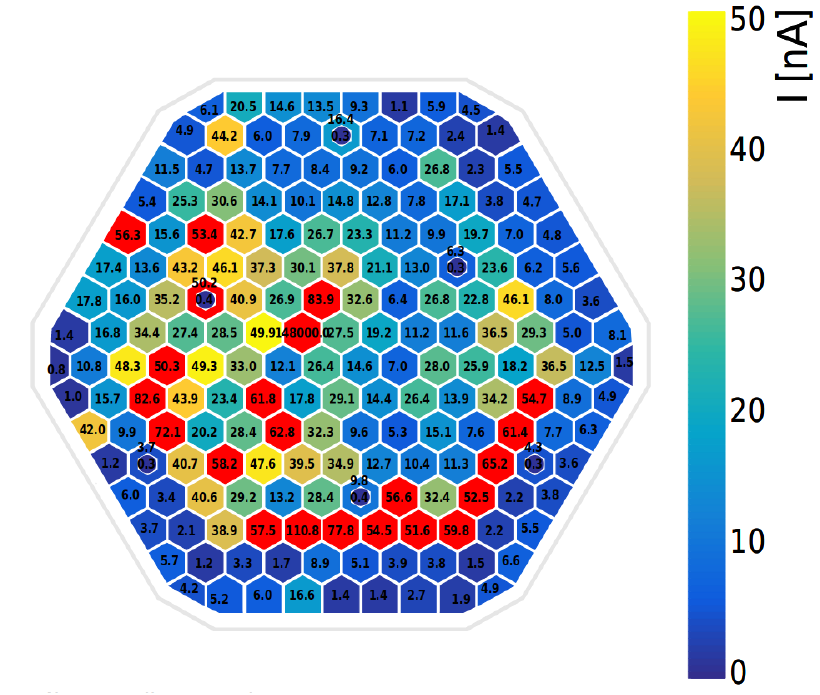} \label{subfig:IV_7needle} }}
    \subfloat[]{{\includegraphics[height=5cm]{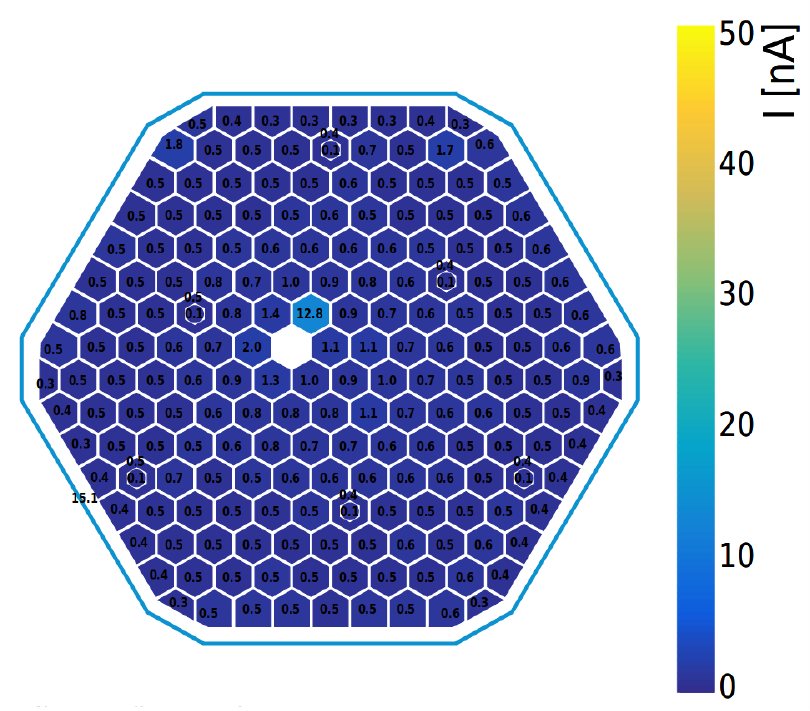} \label{subfig:IV_ARRAY} }}
    \caption{a) Leakage current of an 8-inch low density sensor at a bias voltage of 200V with 7-needle measurement. The red cells have high leakage current. b) The same measurement performed with the ARRAY system.}
    \label{fig:7needle_vs_ARRAY}
\end{figure}

Another possible solution to test single cells would be to use one needle to make electrical contact, keeping all 6 surrounding cells biased with additional 6 needles. Despite being simpler, in some cases the 7-needle configuration can yield significantly higher leakage current when compared with the equivalent measurement from the ARRAY system, as can be seen in Figure \ref{fig:7needle_vs_ARRAY}.

\subsubsection{Thickness Uniformity}

Capacitance measurements can be used to evaluate the thickness uniformity. When using the switch card, the capacitance measured includes both the bulk and the inter-pad capacitances, $C_{SC} = C_{bulk} + C_{int}$.

\begin{figure}[h]
    \centering
    \subfloat[]{{\includegraphics[width=0.45\textwidth]{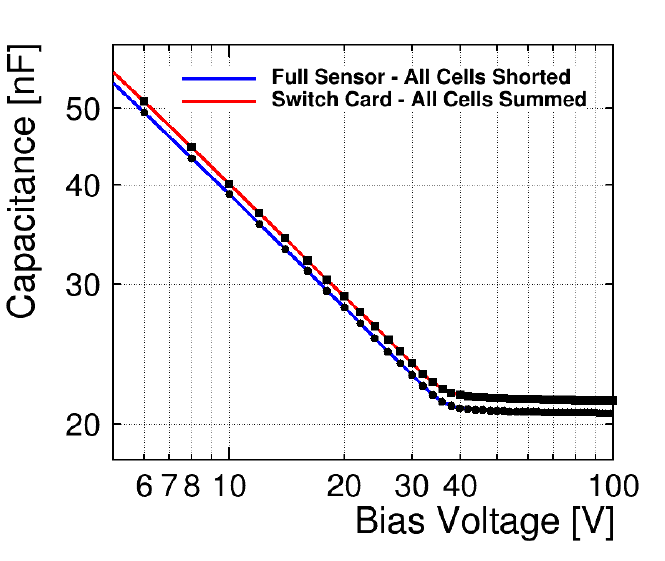} \label{subfig:FS_ARRAY_CV} }}
    \subfloat[]{{\includegraphics[width=0.45\textwidth]{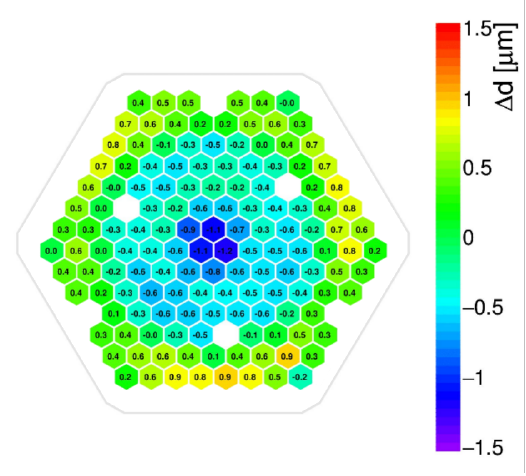} \label{subfig:FS_thicknessUniformity} }}
    \caption{a) CV curves of the full sensor bulk capacitance, measured with all cells shorted; and the sum of the capacitances of all individual cells measured with the switch card. b) The thickness uniformity obtained from the capacitance uniformity. Edge cells and calibration cells are excluded due to not having the same inter-pad capacitance.}
    \label{fig:thickness}
\end{figure}

A CV measurement of the full sensor with all cells shorted was used to eliminate the inter-pad capacitance ($C_{int}=0$). 
The offset between the full sensor CV measurement and the sum of the capacitance of all the cells individually measured by the switch card is attributed to the inter-pad capacitance $C_{int}$, which in the case of the example shown in Figure \ref{subfig:FS_ARRAY_CV} corresponds to $\sim$700 pF.

The capacitance of a fully depleted sensor is well described as a parallel plate capacitor,  $C = \epsilon_{Si} \epsilon_0 A / d$. Where $\epsilon_0$ and $\epsilon_{Si}$ are the vacuum permittivity and the relative permittivity of silicon, respectively; $A$ is the area of the sensor and $d$ is its thickness.
Above the full depletion voltage and assuming that the inter-pad capacitance is the same across the whole sensor, one can assume that variations in the measured capacitance $\Delta C_{SC}$ are due to variations in active thickness, $\Delta d = d_{C} - d_{FS}$ where $d_C$ is the individual cell active thickness and $d_{FS}$ is the active thickness of the full sensor. 
From here follows that $\Delta d = \left(- \epsilon_{Si} \epsilon_0 A \right/ \overline{C}^2_{bulk}) \Delta C_{bulk}$ with $\overline{C}_{bulk}$ being the mean cell bulk capacitance. 
$\Delta d$ can then be calculated for each cell and plotted in a sensor map as shown in Figure \ref{subfig:FS_thicknessUniformity}.

\subsection{Noise Characterization Setup}

The sensor's noise performance is another important quality parameter that provides useful inputs for the optimization of the sensor design during the prototyping phase.
As an example, similar noise measurements performed on strip sensors from the same vendor were found to have different noise behaviour depending on the bulk doping \cite{Canelli2017}. 

A probe card was adapted from the prototype read out electronics used in beam tests \cite{Collaboration:2293646}. 
The only change made to the board was the replacement of the wirebonds by spring loaded pins for making contact.
The readout chip used was the same, the skiroc2CMS, a placeholder for the final chip that is still under development.

To confirm that the measured noise was being driven by the capacitance of the silicon sensor at the input of the skiroc2CMS chip, different voltage steps were used while taking data.

\begin{figure}[h!]
    \centering
    \subfloat[]{{\includegraphics[width=0.33\textwidth]{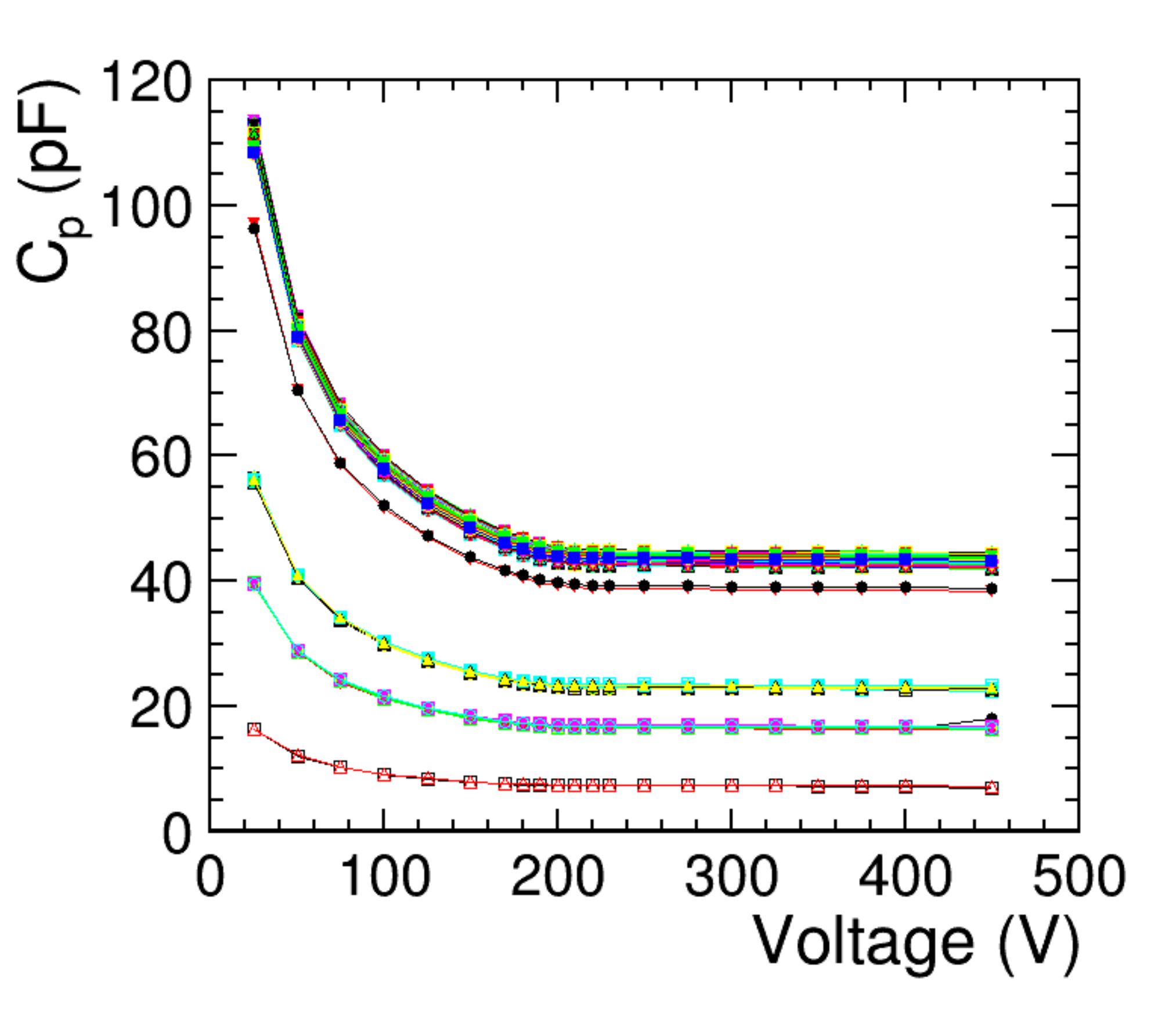} \label{subfig:Noise_CV} }}
    \subfloat[]{{\includegraphics[width=0.33\textwidth]{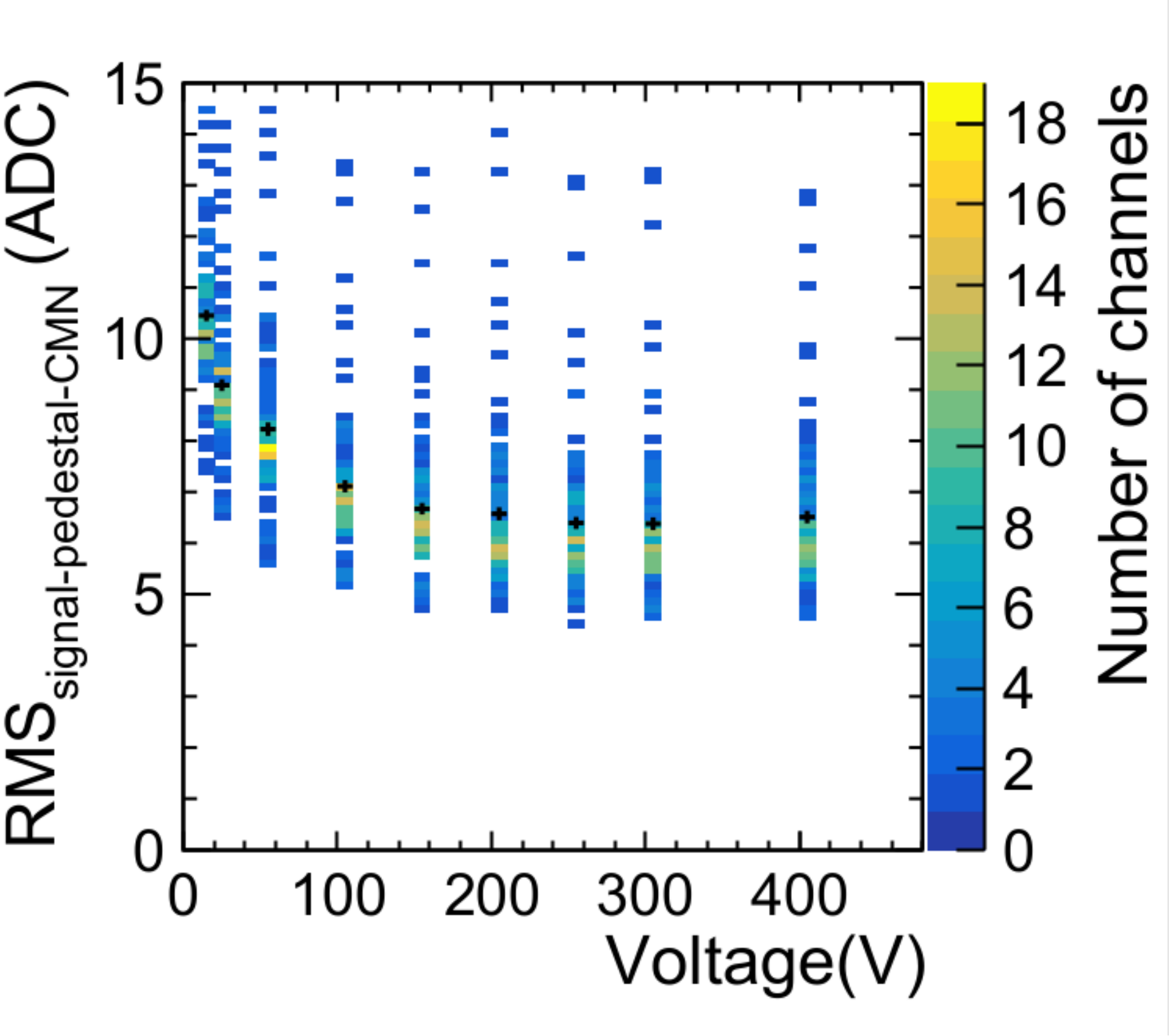} \label{subfig:Noise_Vbias} }}
    \subfloat[]{{\includegraphics[width=0.33\textwidth]{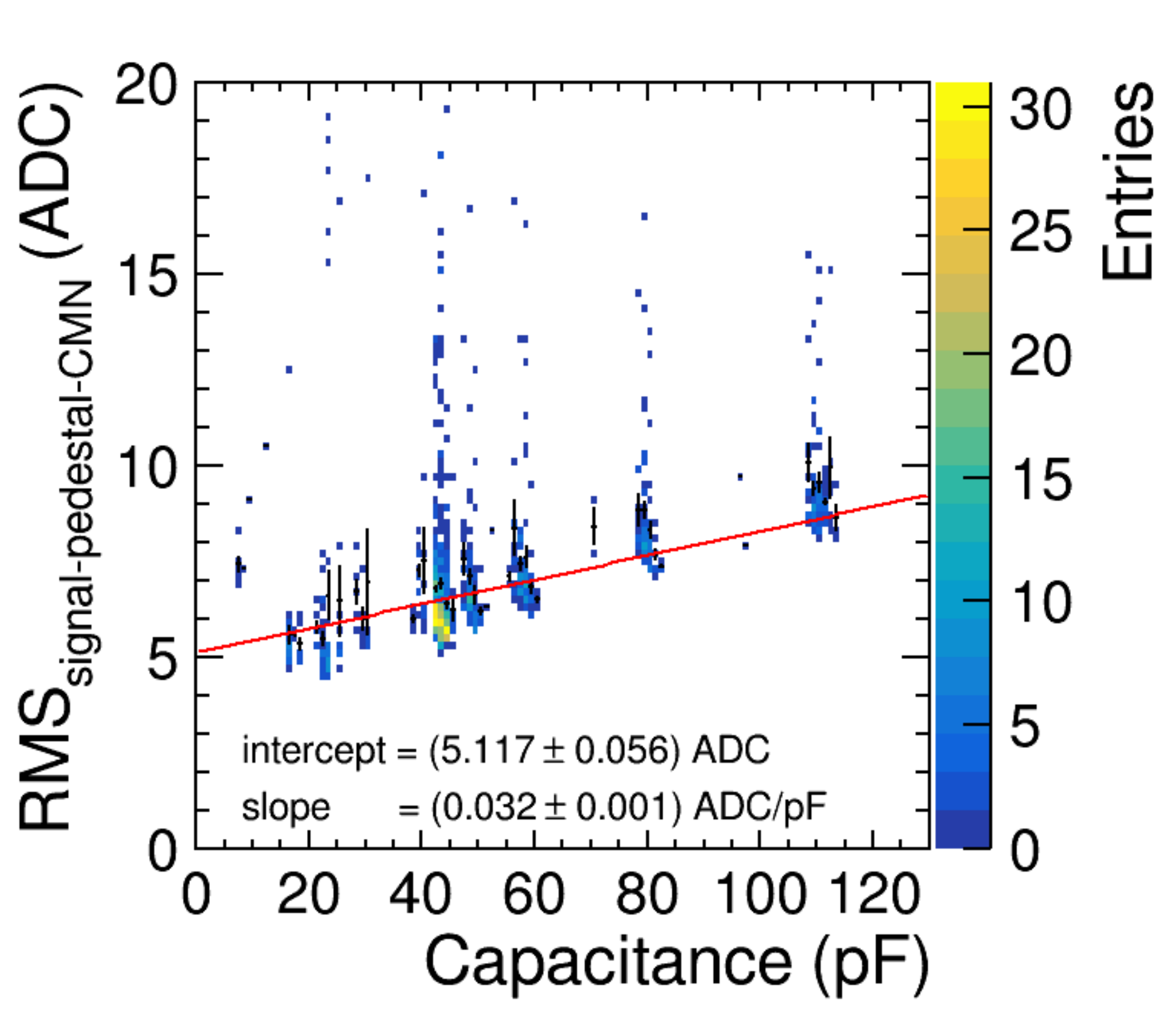} \label{subfig:Noise_Capacitance} }}
    \caption{a) CV curves of the 135 cells of a 6-inch sensor. 5 groups of curves are visible, corresponding to the cells with different areas due to being at the edge of the sensor or because calibration cells. b) The noise measurement of the same sensor follows the same trend as the capacitance. c) The correlation between noise and capacitance is confirmed.}
    \label{fig:Noise}
\end{figure}

The clear correlation between noise and capacitance showed in Figure \ref{fig:Noise} proves that the measured noise results from the sensors' properties. So far, no difference in behaviour was observed between different sensor types.

\subsection{Transient Current Technique (TCT) Setup}

\begin{figure}[h]
    \centering
    \subfloat[]{{\includegraphics[height=6.5cm]{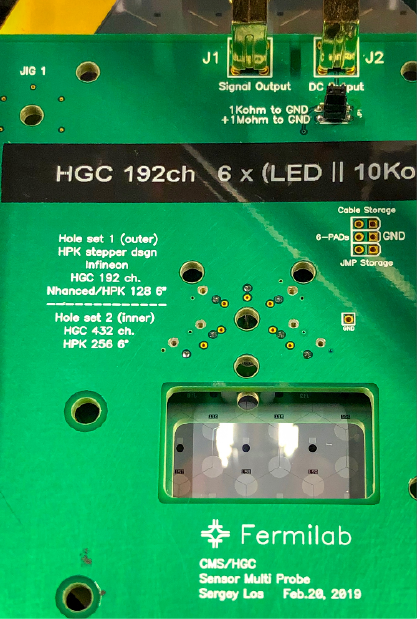} \label{subfig:TCT_picture} }}
    \subfloat[]{{\includegraphics[height=6.5cm]{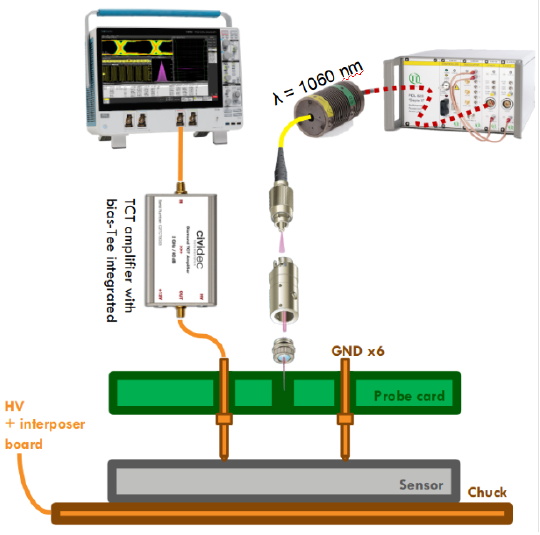} \label{subfig:TCT_schematic} }}
    \caption{a) Photograph of a TCT probe card in contact with a sensor. b) A schematic of the TCT setup.}
    \label{fig:TCT}
\end{figure}

TCT is widely used to measure, among many other things, the charge collection properties of a sensor. Lasers are used to generate charge carriers in the depleted volume of the sensor, resulting in current induction that is then amplified and measured by a fast oscilloscope. A bias-Tee filter is needed in order to separate the AC output signal from the superimposed bias voltage.
A probe card with spring loaded pins, shown in Figure \ref{subfig:TCT_picture}, is used once again to establish electrical contact with the sensor. A schematic of the setup can be seen in Figure \ref{subfig:TCT_schematic}.

In the case of the measurements presented here, an infra-red laser was used, which created a column of electron-hole pairs throughout the thickness of the detector, resulting in a signal similar to that of a minimum ionising particle. The integration of the induced current is a measurement of the collected charge (CC).

\begin{figure}[h]
    \centering
    \subfloat[]{{\includegraphics[width=0.45\textwidth]{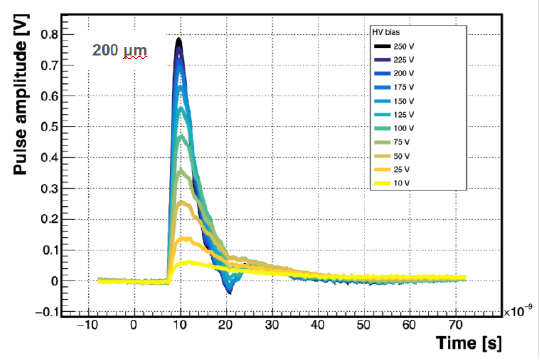} \label{subfig:TCT_waveforms} }}
    \subfloat[]{{\includegraphics[width=0.45\textwidth]{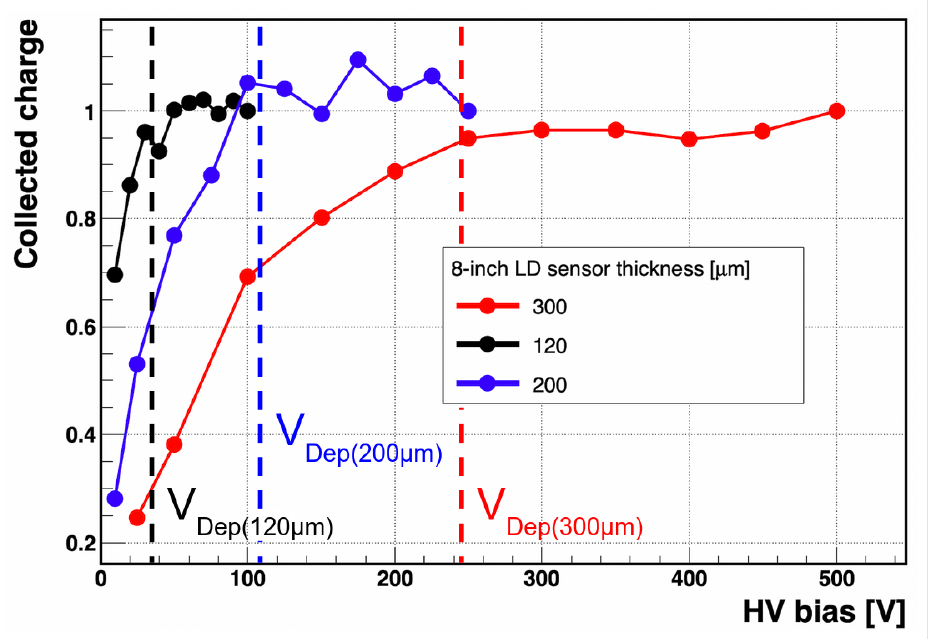}        \label{subfig:TCT_CC} }}
    \caption{a) Raw TCT waveforms acquired by the oscilloscope. b) Normalised charge, obtained from the integration of the waveforms' current, versus the bias voltage for sensors of three different thicknesses. For each thickness, the respective depletion voltage measured by CV is drawn with the dashed line.}
    \label{fig:TCT_results}
\end{figure}

As can be seen in the Figure \ref{subfig:TCT_CC}, there is good agreement between the depletion voltage measured by CV and by CC.

In addition to this configuration, the probe card used for noise measurements can also be used to collect the generated charges. With this configuration one can test charge collection efficiency of the sensor plus the read out electronics.

\newpage

\section{Summary and Conclusions}

Versatile systems for testing large area pad sensors with high granularity were developed. 
The ARRAY system has been performing high quality CV and IV measurements, with a precision better than \SI{0.2}{\pico\farad} and measuring currents as low as \SI{500}{\pico\ampere}.
The probe card with readout electronics has been used successfully to measure the noise of unirradiated sensors and so far no significant difference between the different sensor types was found. 
A TCT setup has been developed and the first charge collection measurements are in agreement with the CV measurements. The methods and infrastructure developed will be used in the assessment of the quality, radiation hardness and noise characteristics of the HGCAL silicon prototype sensors.

\bibliographystyle{plain}
\bibliography{bibliography}

\begin{thebibliography}{1}

\bibitem{Apollinari:2284929}
G.~Apollinari, I.~Béjar~Alonso, O.~Brüning, P.~Fessia, M.~Lamont, L.~Rossi,
  and L.~Tavian.
\newblock {\em {High-Luminosity Large Hadron Collider (HL-LHC): Technical
  Design Report V. 0.1}}.
\newblock CERN, Geneva, 2017.

\bibitem{Canelli2017}
Florencia Canelli, Benjamin Kilminster, Thea Aarestad, Lea Caminada, Annapaoloa
  {De Cosa}, Riccardo {Del Burgo}, Silvio Donato, Camilla Galloni, Andreas
  Hinzmann, Tomas Hreus, Jennifer Ngadiuba, Deborah Pinna, Giorgia Rauco, Peter
  Robmann, Daniel Salerno, Korbinian Schweiger, Claudia Seitz, Yuta Takahashi,
  and Alberto Zucchetta.
\newblock {P-type silicon strip sensors for the new CMS tracker at HL-LHC}.
\newblock {\em Journal of Instrumentation}, 12(06):06018, 2017.

\bibitem{Pitters:2668752}
Florian Pitters, Erica Brondolin, Dominik Dannheim, Szymon Kulis,
  Andreas~Alexander Maier, Thorben Quast, and Eva Sicking.
\newblock {ARRAY: An Open Source, Modular and Probe-Card based System with
  Integrated Switching Matrix for Characterisation of Large Area Silicon Pad
  Sensors}.
\newblock {\em Nucl. Instrum. Methods Phys. Res., A},
  940(arXiv:1903.10262):168--173. 6 p, Mar 2019.

\bibitem{Collaboration:2293646}
\text{CMS Collaboration}.
\newblock {The Phase-2 Upgrade of the CMS Endcap Calorimeter}.
\newblock Technical Report CERN-LHCC-2017-023. CMS-TDR-019, CERN, Geneva, Nov
  2017.

\end{thebibliography}

\end{document}